# CHAINPLOT: A Teaching Tool to Demonstrate Vibrations of Atoms on a Chain and the Cautionary Tale of a Fundamental Error Propagated Over the Last 63 Years.

A.M. Glazer, Clarendon Laboratory, Parks Road, Oxford OX1 3PU, UK & Department of Physics, University of Warwick, Coventry CV4 7AL

**Abstract**

Simulation of the so-called monatomic and diatomic chains by the computer program CHAINPLOT is described. The simulation provides useful teaching material for undergraduate condensed matter physics lecture courses, and has revealed that for the last 63 years a fundamental description of the actual atomic motions has been copied in in many textbooks, despite being incorrect or at least misleading.

**Background**

One of the standard subjects taught in most condensed matter courses is that of the vibrations of atoms on a chain. Almost all students of the crystalline state will at some time have seen the derivation of the so-called monatomic chain, in which identical atoms are placed in a 1-dimensional periodic array with a fixed, usually nearest-neighbour, force constant. This is often followed by the diatomic chain in which alternating atoms of different mass are used, sometimes with alternating force constants, to create the so-called acoustic and optic branches (see for example (Kittel, 1953; Levy, 1968; Ashcroft & Mermin, 1976; Blakemore, 1985; Pain, 2000; Dove, 2003)). The justification for teaching this rather artificial model in a Condensed Matter Physics (CMP) course is that it gives the student an introduction to concepts such as normal modes, phonons and Brillouin Zones. However, many of the treatments in the standard texts are superficial and tend to gloss over some important ideas that underlie a true understanding of the solid state. The most important concept underlying CMP is that of translational symmetry and its description by a *lattice*. Unfortunately, despite its simplicity and the fact that it is defined by the International Union of Crystallography (Burns & Glazer, 2013; Aroyo, 2016), experience has shown that students tend to confuse the concepts of *lattice* and *structure*. The following is based on my own experience in teaching CMP to undergraduates for over 45 years.

Some time ago it occurred to me that it would be useful for pedagogical reasons to write a computer program to simulate the motions of the atoms in such chains, especially since, surprisingly, there did not seem to be any useful simulations of this type available on the internet. As we shall see below, there are aspects of this topic that are revealed more obviously when the software is run, and these give the student a better feel for the ideas behind normal modes in solids.

It is worth revisiting the standard treatment, which begins with consideration of wave propagation in a one-dimensional linear chain of atoms (Figure 1(b)), which is in fact a representation of a 1D crystal structure with single atoms separated by a distance $a_1$. In order to emphasize the importance of translational symmetry, I have added in Figure 1(a) the associated lattice with the repeat distance $a_1$ marking *primitive* unit cells i.e. unit cells with one lattice point. At first sight, the two diagrams look similar, since there is just one atom per primitive unit cell. This happens too when considering real three-dimensional crystals since many of the structures studied at this level are simple e.g. copper or molybdenum, and have only one atom per lattice point: from experience this often causes misunderstanding over the distinction between a lattice point and an atom. These waves can be transverse (in which case there are two such transverse, degenerate components) or longitudinal.

In order to determine the characteristics of this simple system we need to determine the classical equations of motion for the atoms. In its simplest form, this is done by using Hook's Law and thinking of the bonds between the atoms as if they were springs with force constant $C$. As one atom moves, it influences its neighbours, next-neighbours and so on causing them to move in sympathy. It is this that creates the wave. It is then a simple matter of solving the equations of motion. Usually in undergraduate physics courses it is sufficient to use the harmonic approximation with nearest-neighbour force constants, but it is relatively easy to incorporate higher-order forces if required.

We start with displacements of the atoms denoted by $u_s$, the displacement of the s'th atom in the chain (Figure 1b).

From Hook's Law the equation of motion is then given by

$$M\ddot{u}_s = C(u_{s+1} - u_s) + C((u_{s-1} - u_s) \tag{1}$$

If we now assume cyclic boundary conditions, a nearest-neighbour wave-like solution can be assumed:

$$u_s = U e^{i(ka_1 - \omega t)} \tag{2}$$

We then get

$$\begin{aligned} -M\omega^2 &= C\left(e^{ika_1} + e^{-ika_1} - 2\right) \\ &= 2C(\cos ka_1 - 1) \end{aligned} \tag{3}$$

Therefore, we find that the angular frequency is

$$\omega = \sqrt{\frac{4C}{M}} \sin \frac{ka_1}{2} = \omega_{max} \sin \frac{ka_1}{2} \tag{4}$$

From this one can calculate, for instance, a travelling wave group velocity

$$v_g = \frac{d\omega}{dk} \tag{5}$$

which for $k = \pi/a_1$ is equal to 0. This value of $k$ then marks the so-called Brillouin Zone boundary, which because of it translational equivalence to $k = -\pi/a_1$, which means that this wave travels in the opposite direction with opposite phase, thus creating a stationary wave. For wave-vectors away from the zone boundary, travelling waves occur with a non-zero group velocity. Here is an example of where some books tend to gloss over an important point, thus leading to possible confusion. The use of cyclic boundary conditions means that the wave solutions should be stationary, so how can there be travelling wave solutions? The answer is that the Brillouin Zone actually spans the whole range given by $-\pi/a_1 <= k <= \pi/a_1$, so that for every travelling wave to the right there is an equivalent antiphase wave travelling in the opposite direction, together creating stationary wave solutions. Note, however, often the reduced Brillouin Zone scheme is drawn in the literature in the region only from $k = 0$ to $\pi/a_1$, one half of the Brillouin Zone, and this just adds to the misperception.

In the diatomic case, there are atoms of different mass alternating along a chain (Figure 2). This time note that the translational invariance (Figure 2(a)) is given by a series of lattice points $a_2 = 2a_1$. In this case, there are two atoms per primitive unit cell. Some books (e.g. Kittel, 1953) prefer to use the nearest-neighbour distance $a_1$ in the treatment. Now, while this is not incorrect, the danger of this is that it conceals the importance of the fundamental notion of translational symmetry and the true meaning of the Brillouin Zone. Furthermore, it leads to the Brillouin Zone boundary being set at $k = \pm 2\pi/a_1$ with, confusingly as far as students are concerned, reciprocal lattice points at 0, $\pm 4\pi/a_1, \pm 8\pi/a_1$ rather than expected sequence of $k = 0, \pm 2\pi/a_2, \pm 4\pi/a_2,...$ and Brillouin Zone boundary at $k = \pm \pi/a_1$. The student needs to understand, here, that the Wigner-Seitz construction of the Brillouin Zone in fact gives a primitive cell in $k$-space, and this needs to be compared with a primitive cell in direct space. In this context, recall that the number of wave states in the Brillouin Zone is equal to the number of primitive unit cells in direct space. The recent book by Simon (Simon, 2013) uses the lattice repeat as in this paper.

The procedure starts by writing the equations of motion for nearest-neighbour atoms of masses $M_1$ and $M_2$ in terms of a Hooke's Law formula with force constant $C$ and displacements $u$, the subscript $s$ indicating atomic position in the chain:

$$M_1 \ddot{u}_s = C[u_{s+1} - u_s] + C[u_{s-1} - u_s]$$
$$M_2 \ddot{u}_{s+1} = C[u_{s+2} - u_{s+1}] + C[u_s - u_{s+1}] \quad (6)$$

The solution is then given by assuming two wave functions with angular frequency $\omega$:

$$u_s = U e^{i[ksa_2/2 - \omega t]}$$
$$u_{s+1} = V e^{i[k(s+1)a_2/2 - \omega t]} \quad (7)$$

The resulting frequency dispersion equation is then given by

$$\omega^2 = \frac{C(M_1 + M_2)}{M_1 M_2} \pm C\left(\left[\frac{(M_1 + M_2)}{M_1 M_2}\right]^2 - \frac{4}{M_1 M_2} \sin^2 ka_2/2\right)^{\frac{1}{2}} \quad (8)$$

For a wave-vector at the Brillouin Zone boundary, there are two solutions with a frequency gap due to the different masses:

$$\omega_1 = \sqrt{\frac{2C}{M_1}}$$
$$\omega_2 = \sqrt{\frac{2C}{M_2}} \quad (9)$$

showing that in one solution only the heavy atoms move while in the other only the light atoms move.

It can also be shown that the ratio of the amplitudes U to V is given by

$$\frac{U}{V} = \frac{2C \cos ka/2}{2C - M_1 \omega^2} = \frac{2C - M_2 \omega^2}{2C \cos ka/2} \quad (10)$$

For $k = 0$

$$\omega = 0 \quad \text{acoustic branch}$$
$$\omega = \sqrt{2C\left(\frac{1}{M_1} + \frac{1}{M_2}\right)} \quad \text{optic branch} \quad (11)$$

Then, substituting into Equation(10), we find that

$$\frac{U}{V}=1 \qquad \text{acoustic branch}$$

$$\frac{U}{V}=-\frac{M_1}{M_2} \quad \text{optic branch}$$

(12)

In other words both atoms have the same amplitude for the acoustic mode but different and opposite amplitudes for the optic mode. However, for any intermediate value of $k$ inside the Brillouin Zone boundary, Equation (10) shows that the amplitudes on the acoustic and optic modes must be different.

### Computer program

CHAINPLOT has been successfully used for several years as part of the 3$^{rd}$-year undergraduate course on Condensed Matter Physics at Oxford University. It was written using Delphi6 in Pascal and can be downloaded freely from http://www.amg122.com/programs/chains.html. There are no restrictions on its use.

Figure 3 shows an example of the computer screen for the monatomic case with a wave-vector intermediate between $k = 0$ and $\pi/A$, where A is the lattice repeat distance. The screen also depicts (bottom right) the $\omega$-$k$ dispersion curve with the corresponding wave-vector indicated. Controls (top right) are given to change the speed of the plot, to select the wave-vector and to introduce damping, if required. In addition, the user can choose to display either transverse or longitudinal vibrations. A next-nearest neighbour model is also available. Figure 4 shows examples of the generated waves for different values of the wave-vector. In teaching this subject, it is important to explain that one cannot have a wave with wavelength less than 2A (i.e with wave-vector $k > \pi/A$ outside the Brillouin one boundary). A nice feature of the program is that one can choose to plot such an unphysical wave, showing that this is equivalent to a real wave travelling in the opposite direction. This is shown by the dotted curves. The bottom right diagram shows what happens for $k = \pi/A$, where the two waves travelling in opposite directions have the same wavelength and amplitude.

Figure 5 shows the screen for the diatomic case in which two different masses ($M1 = 8$, $M2 = 4$, arbitrary units) have been selected. Note that the distance **a = 2A** is the true repeat distance between *equivalent* atoms i.e. the lattice repeat distance. In this case a wavelength of **2.3a** is shown. A facility to choose between acoustic and optic branches is also available. Note that the dispersion curve plot at bottom right now shows the frequency gap at the Brillouin Zone boundary. By changing the masses of the atoms to become more equal, the student can then explore the effect this has on the gap and how the diatomic chain morphs into the monatomic chain when the masses are equal. In addition, one can again include the effect of choosing a *k*-vector lying outside the Brillouin zone. In the Figure this wave is indicated by dotted lines and can be observed to form a wave travelling in the opposite direction. Another button allows for alternating force constants while another incorporates mass differences as well.

**Comment**

After running this program, I was surprised to see that in the diatomic chain the heavy and light atoms were displaced on two different waves, both with the same wavelength and phase. Often one reads that in the acoustic branch the atoms move together while in the optic branch they move opposite to one another. But this statement fails to add that the *amount* that they move is different between heavy and light atoms. However, this must be so, because for any value of *k* different from the zone-boundary value, the solution of the wave equations for, say, $0 <= k < \pi/A$, gives rise to *running* waves travelling from left to right, and not *standing* waves. Therefore, as the wave travels in one direction the maximum amplitudes of the heavy and light atoms *cannot* be the same. This can also be seen by thinking about what happens for a wave-vector near the zone boundary value. At the zone boundary one set of atoms does not move while the other one does, and so it is obvious that for *k* just less than this the two types of atom must exhibit different amplitudes, as found in Equation (10) .

Yet, this is not how it is portrayed in many text books and on the internet. One finds in most solid state books, for example (Kittel, 1953; Levy, 1968; Blakemore, 1985; Pain, 2000; Dove, 2003) a diagram similar to Figure 6, where it can be seen that both heavy and light atoms have been drawn to lie on the same sinusoidal wave. It is especially obvious for the acoustic branch. Similarly see http://www.chembio.uoguelph.ca/educmat/chm729/phonons/optmovie.htm, http://slideplayer.com/slide/6258043/ . Certainly for low values of *k*, they do look at first sight to be on the same wave but on increasing *k* the difference becomes clear.

I have searched to see where and when this fundamental error first appeared. The books by Brillouin and by Wannier have it correct (Brillouin, 1946; Wannier, 1959), but as far as I can tell, this diagram first appears in the 1st Edition of Kittel's classic textbook ((Kittel, 1953) and in every edition since. The problem arises because Kittel showed that for small values of the wave-vector *k* one obtains the result U = V: however, this is just an approximation, being strictly true only for *k* = 0, and so to draw the acoustic motion as in Figure 6 is at best misleading. Kittel's drawing for the optic mode case seems to indicate that the heavy atoms are displaced in equal and opposite directions with the same amplitude, yet he shows that $U/V = -M_1/M_2$ for small *k*. It appears that this is a classic case where this diagram has simply been widely copied for the last 63 years without thinking it through.

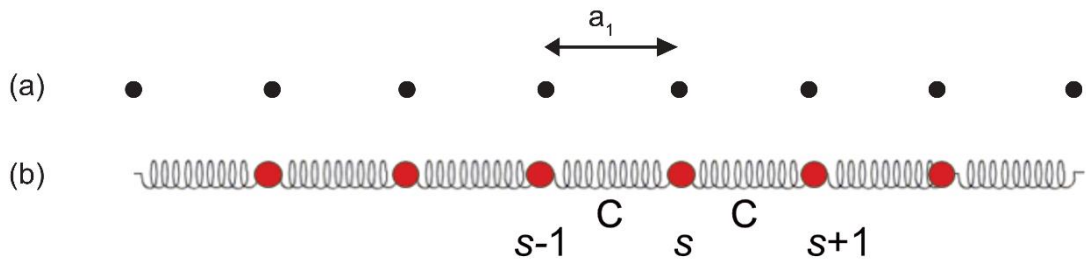

**Figure 1.** (a) Primitive lattice of spacing $a_1$. (b) a linear chain of atoms of mass $M$ with repeat distance $a_1$

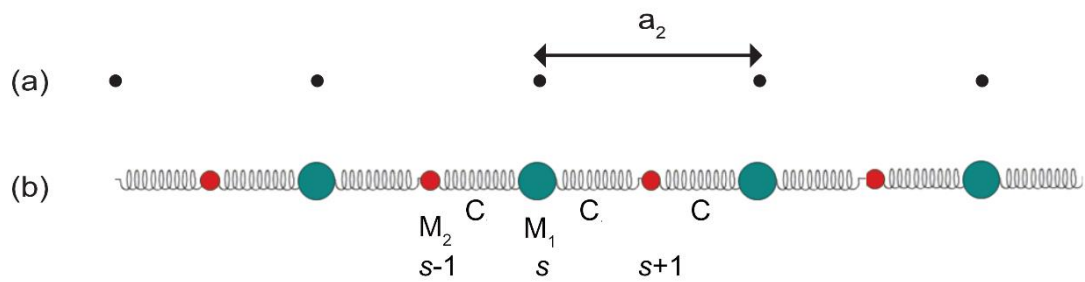

**Figure 2** (a) Primitive lattice of spacing $a_2 = 2a_1$ (b) The diatomic chain with alternating atoms of different mass $M_1$ and $M_2$ but with equal force constants $C$ between them. Note that the distance $a_2$ is the repeat distance, not the distance between neighbouring atoms.

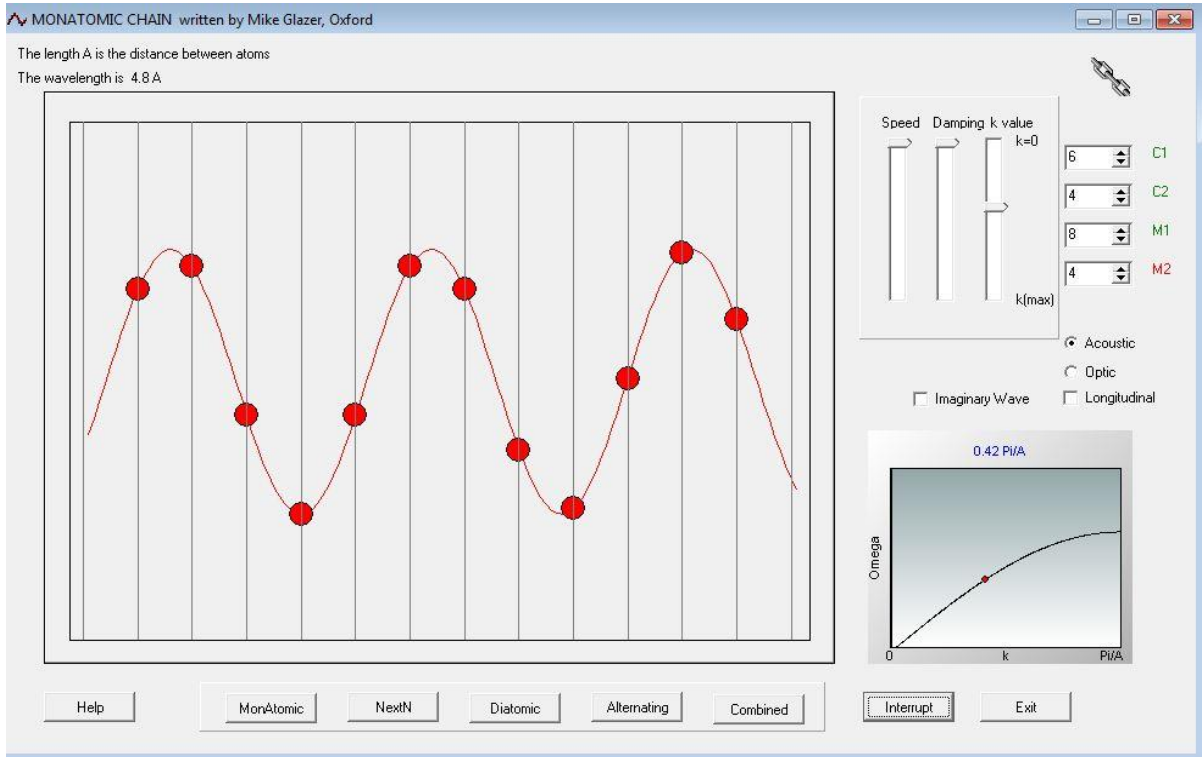

**Figure 3 Screen plot for a transverse wave on a monatomic chain from the CHAINPLOT program. The wavelength shown is 4.8A (k = π/2.4A).**

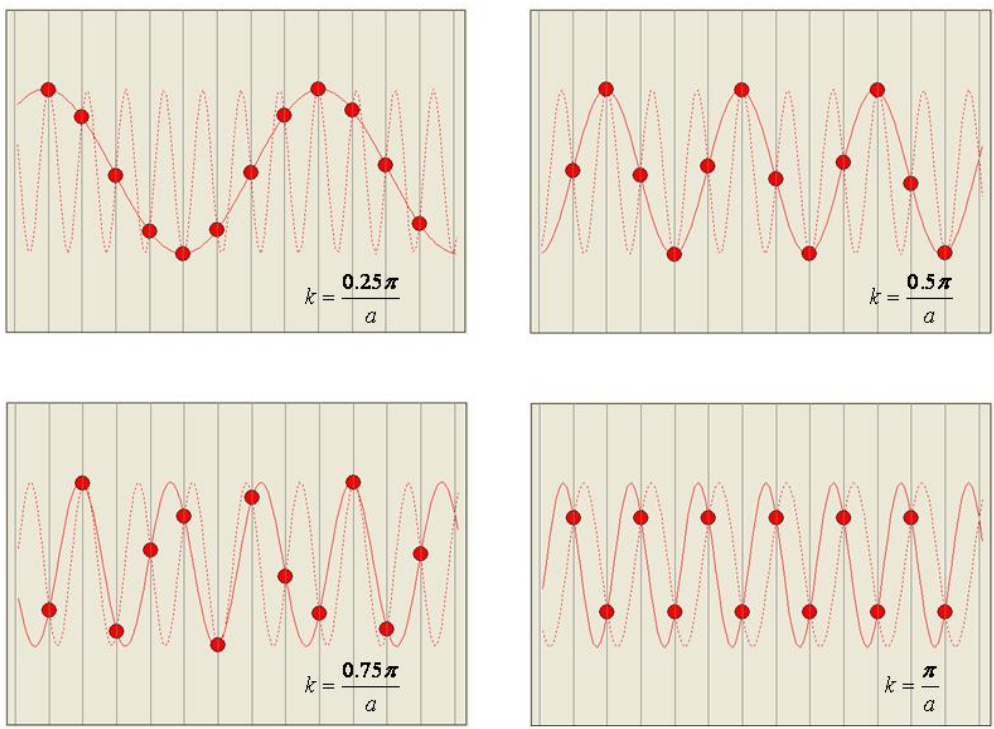

**Figure 4 Examples of transverse oscillations in a monatomic chain for different values of wave-vector.**

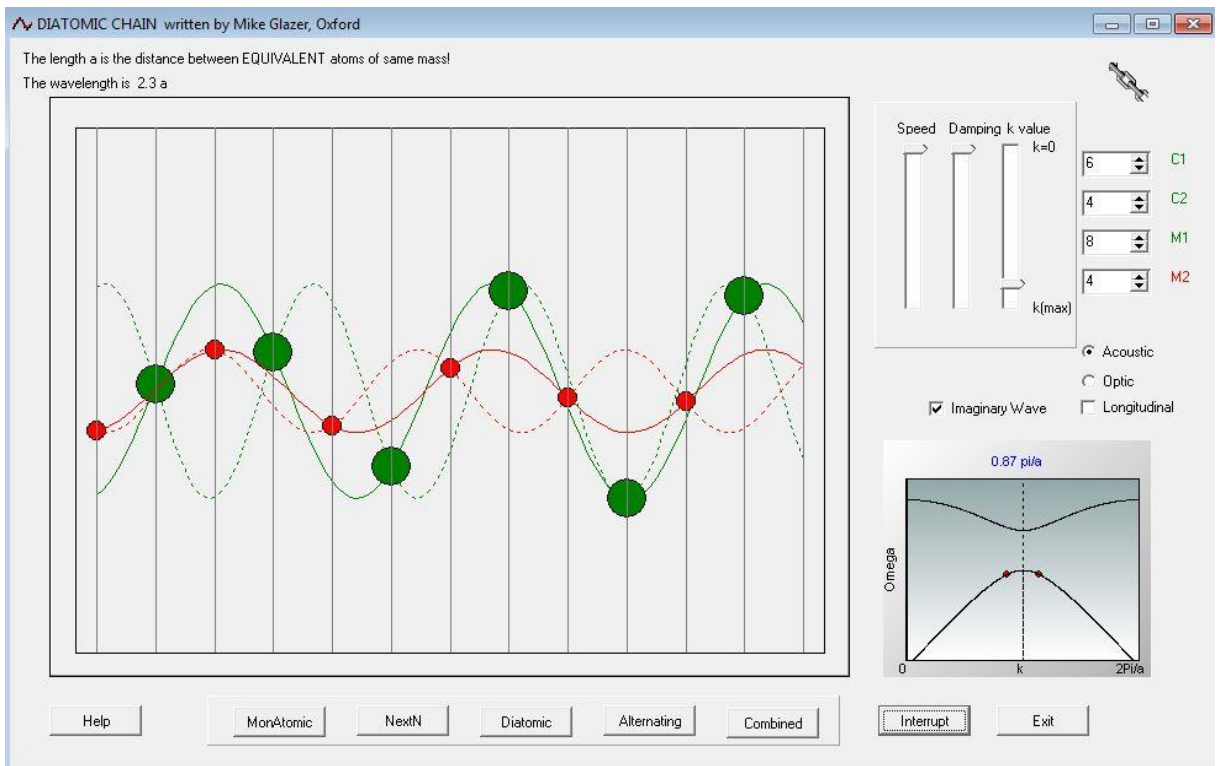

**Figure 5 Screen plot for a transverse acoustic wave on a diatomic chain from the CHAINPLOT program. The wavelength shown is 2.3a (k = π/1.15a). The green curve is for heavy atoms while the red curve is for light atoms. The dashed curves are for a wave-vector outside the first Brillouin zone.**

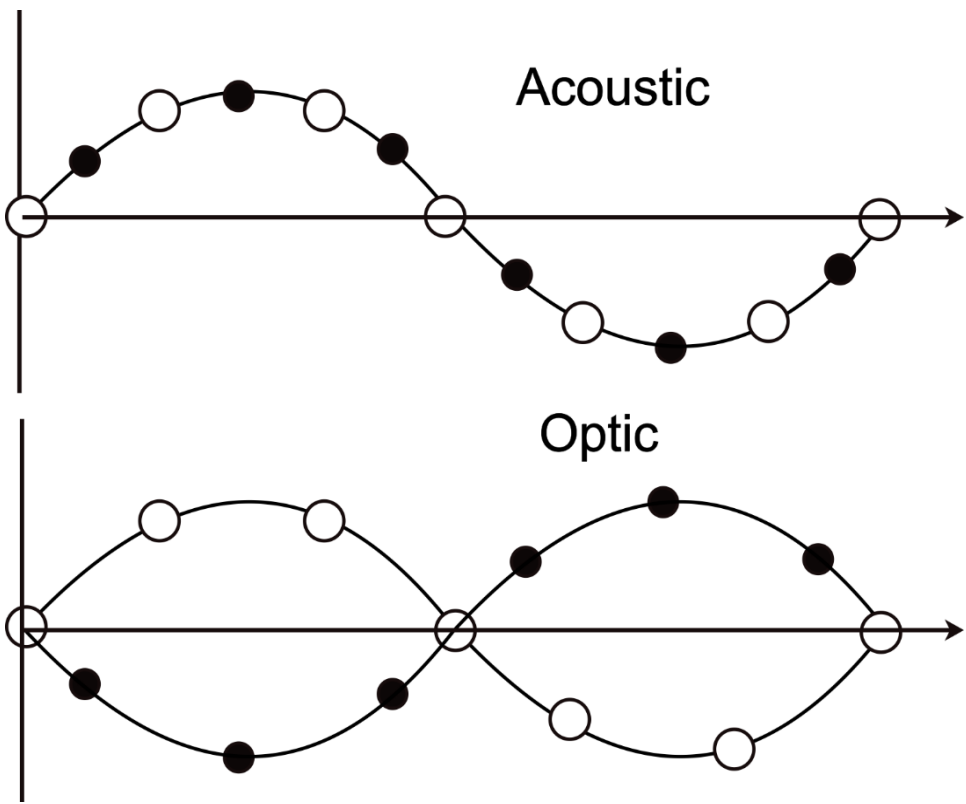

**Figure 6 Incorrect portrayal of acoustic and optic modes with $0 < k < k_{ZB}$, adapted from Figure 4.4 of (Kittel, 1953).**